\pdfoutput=1
\documentclass{article}
\usepackage{graphicx}

\usepackage[final,dblblindworkshop]{neurips_2026}   

\workshoptitle{Machine Learning and the Physical Sciences}

\usepackage[normalem]{ulem}
\usepackage{wrapfig}  
\usepackage[utf8]{inputenc}
\usepackage[T1]{fontenc}
\usepackage{url}
\usepackage{booktabs}
\usepackage{amsfonts}
\usepackage{amssymb}
\usepackage{microtype}
\usepackage{xcolor}
\usepackage{amsmath}
\usepackage{bm, color, colortbl}
\definecolor{linkcolor}{rgb}{0.6,0,0}
\definecolor{citecolor}{rgb}{0,0,0.75}
\definecolor{urlcolor}{rgb}{0.12,0.46,0.7}
\usepackage[citecolor=citecolor]{hyperref}
\hypersetup{colorlinks,linkcolor=linkcolor,citecolor=citecolor,urlcolor=urlcolor}

\definecolor{newcolor}{rgb}{0.80,0.33,0.0}

\definecolor{editcolor}{rgb}{0.0,0.50,0.30}  

\definecolor{cutcolor}{rgb}{0.45,0.45,0.45}  

\definecolor{vfivecolor}{rgb}{0.0,0.25,0.85}   

\definecolor{camrcolor}{rgb}{0.80,0.0,0.60}

\title{Wavelet-CFM for the extragalactic millimeter-wave sky}

\author{%
  M.~Mebratu$^{a,b,c}$, \quad W.~L.~K.~Wu$^{d,b,c}$, \quad S.~Gasiorowski$^{b}$ \\
  $^{a}$Stanford University Department of Physics, Stanford, CA 94305, USA \\
  $^{b}$SLAC National Accelerator Laboratory, Menlo Park, CA 94306, USA \\
  $^{c}$Kavli Institute for Particle Astrophysics and Cosmology, Stanford, CA 94305, USA \\
  $^{d}$California Institute of Technology, Pasadena, CA 91125, USA \\
  \texttt{mmebrat1@stanford.edu}, \texttt{wlw@caltech.edu}, \texttt{sgaz@slac.stanford.edu}}

\begin{document}

\maketitle

\begin{abstract}
We develop a Wavelet-factorized Conditional Flow Matching (Wavelet-CFM) framework to generate fast, accurate simulations of correlated astrophysical fields including gravitational lensing convergence ($\kappa$), the kinetic and thermal Sunyaev-Zel'dovich (kSZ and tSZ) effects, and cosmic infrared background (CIB) emission. Together these make up every bright extragalactic component of the millimeter-wave (mm-wave) sky apart from radio sources.
Our approach combines multi-scale wavelet decomposition with conditional flow matching models, trained on $\sim$9$\times$9 deg$^2$ patches at 1-arcminute resolution.
The four fields are generated jointly, including their cross couplings, and the model has a tractable likelihood, which to our knowledge has not previously been achieved for this set of components.
The power spectra, one-point distributions and Minkowski functionals of the generated maps match a validation set to a few percent.
These results show promise for enabling
field-level cosmological analyses of mm-wave data.
\end{abstract}

\section{Introduction}
Precision cosmology from cosmic microwave background (CMB) observations depends  on accurate modeling of foregrounds and secondary anisotropies that contaminate the primordial signal~\citep{cibtszradio, vanengelenfgbias, radiocibimpactontszandlensing, plank2018, actdr6foregrounds}. These effects arise from structure formation processes occurring after recombination and include gravitational deflection of CMB photons, interactions with hot gas, and infrared emissions from dusty galaxies~\citep{tszoriginal, kszoriginal, cibdetection,  cibreview, dezotti}.

Traditional cosmological analyses compress observed maps into summary statistics such as power spectra.
In contrast, field-level inference directly analyzes pixel-level data, accessing non-Gaussian features and morphological properties lost in conventional approaches~\citep{10.1093/mnrasl/slab081, mancini2024fieldlevelcosmologicalmodelselection}. However, their practical application is limited by computational demands. Foreground simulations required for field-level inference are expensive to generate due to their non-Gaussian statistics and complex multi-scale correlations between different astrophysical components~\citep{websky, yuuki, halfdome}.

To address these computational challenges, we develop a Wavelet-factorized Conditional Flow Matching framework for the extragalactic millimeter-wave sky. Our model jointly generates the gravitational lensing convergence ($\kappa$), the kinetic and thermal Sunyaev-Zel'dovich (kSZ and tSZ) effects and the cosmic infrared background (CIB), which together make up every bright extragalactic component of the sky at these frequencies apart from radio sources, which are routinely masked in analyses of this kind. The fields are generated with their cross couplings and the model has a tractable likelihood, which the existing models of this sky do not offer.\footnote{\citet{mmDL} generate all five bright components, including radio sources, but their model does not provide a tractable likelihood.}

\section{Methods}
\label{sec:methods}
\textbf{\textit{Conditional flow matching models}}
Flow matching (FM)~\citep{lipman2023flow, DBLP:conf/iclr/LiuG023, 2022arXiv220915571A, tong2024improving} learns to generate data by parameterizing the vector field of an ordinary differential equation (ODE) that transports samples $x$ from a simple prior $p_0$ at $t=0$ to the target data distribution $p_1$ at $t=1$. Conditional Flow Matching (CFM)~\citep{lipman2023flow, tong2024improving} makes this tractable by regressing on a vector field defined conditionally on a pair of endpoints $(x_0, x_1) \sim \pi(x_0, x_1)$, drawn here from the independent coupling $\pi(x_0,x_1) = p_0(x_0)p_1(x_1)$. We use the straight probability path $x_t = (1-t)x_0 + t x_1$, for which the conditional vector field is simply $u_t(x \mid x_0, x_1) = x_1 - x_0$. The objective learns a neural network $v_\theta$ that approximates this field:
\begin{equation}
\label{eq:loss}
\mathcal{L}_{CFM}(\theta) = \mathbb{E}_{t\sim\mathcal{U}[0,1],(x_0,x_1)\sim\pi,x_t\sim p_{t|0,1}(\cdot|x_0,x_1)}\left[\|v_\theta(x_t, x_0, x_1) - u_t(x_t | x_0, x_1)\|^2\right]
\end{equation}

Once trained, we generate samples $x_1$ by integrating the learned vector field $v_\theta$ from $t=0$ to $t=1$ using an ODE solver. Integrating the divergence of $v_\theta$ along the same trajectory gives the log-likelihood of any map under the model, which is what makes the model usable inside an inference pipeline rather than only as a sample generator. We estimate the divergence with the Hutchinson trace estimator~\citep{2018FFJORD, Hutchinson1989}. Likelihoods computed with our model show good agreement on analytic Gaussian maps, for which the density is known exactly (Appendix~\ref{app:gauss}).

\textbf{\textit{Wavelet-CFM}}
extends the CFM framework with multi-scale wavelet decomposition, replacing the normalizing flows~\citep{NflowDinh2014NICENI, Nflowreview, NflowDinh2016DensityEU, NflowJimenezRezende2015VariationalIW, NflowKingma2018GlowGF, Nflowadam, Nflowternf, NflowH1flow, NflowStachurski2023CosmologicalIU} in Wavelet Flow~\citep{waveletflowCS, waveletflowCosmology, WaveletValiuddin2022EfficientOD, Mebratu:2025vxc} with CFMs. 
In this framework, we employ Discrete Wavelet Transforms (DWTs) to decompose an input map $x \in \mathbb{R}^{N \times N \times C}$ into multiple scales. At each level $i$, the map is split into low-frequency coefficients $x_{i+1}$ through a low-pass filter and high-frequency detail coefficients $x^d_{i+1}$ via a high-pass filter. This decomposition is performed recursively with $i$ ranging from 1 to $k$. We use Haar wavelets for all wavelet transforms in this work.

For levels $i < k$, we train a CFM to learn the conditional distribution $p(x^d_i | x_i)$
and for level $k$ we train a CFM to learn the distribution $p(x_k)$.
The overall distribution factorizes as:
$p(x) = p(x_k) \prod_{i=1}^k p(x^d_i | x_i)$.
Training proceeds independently for each scale, enabling efficient parallelization. Recent wavelet-domain flow matching models~\citep{wfmaccarino2026waveletflowmatchingmultiscale, tur2026wfm, wfmDanese_2026} instead learn a single network jointly over all sub-bands. Our factorization follows Wavelet Score-Based Generative Models~\citep{guth2022waveletscorebasedgenerativemodeling}, which train a separate score-based model at each level, and to our knowledge this per-level factorization with CFM is novel. During inference we generate samples progressively from coarsest to finest scale: first sampling $x_k$ from the base distribution, then, at each level $i$, generating detail coefficients $x^d_i$ conditioned on the low frequency component $x_{i}$ and reconstructing $x_{i+1}$ via inverse wavelet transforms.

\textbf{\textit{Hybrid and condition-dependent priors}}
Rather than using a white noise prior ($p_0 = \mathcal{N}(0, I)$) uniformly across all scales, we adopt the Hybrid Component Correlated (HCC) approach of~\citep{Mebratu:2025vxc}: white noise at the coarsest scale, and at the detail levels a Component Correlated (CC) prior derived from the power spectra of and the cross-spectra between the component maps in wavelet-transformed space. This maintains training stability at large scales while letting the prior carry the correlations between the fields at the finer ones.
At the detail levels we also let the prior depend on the low-frequency map the flow is
conditioned on, in the spirit of the condition-dependent source distributions of~\citep{cdprior}. For each field and wavelet band we tabulate, over the training set, the mean squared
detail coefficient in twenty quantile bins of the low-frequency pixel value. We then divide by its average
over all pixels and multiply each source pixel by the square root of the entry its low-frequency
pixel falls in. The prior then carries the amplitude of the detail coefficients, which is far larger
around a cluster than in an empty region, and the network only has to model their shape.

\textbf{\textit{Training Data}}\label{sec:td}
We use the \texttt{Agora} extragalactic foreground simulations~\citep{yuuki} to construct our training dataset. These simulations provide realistic maps of multiple correlated CMB secondaries and foregrounds.

Our training samples consist of 512$\times$512 pixel cutouts extracted from full-sky maps at 150~GHz via \texttt{pixell},\footnote{https://github.com/simonsobs/pixell} with 1.0-arcminute resolution covering approximately 9$\times$9 deg$^2$ patches.
Before any transformation we mask the brightest tSZ clusters, which are detected with a matched filter over core radii $\theta_c \in [0.25, 10]$ arcmin and masked out to $3\theta_c$. The masked regions, on average about $4\%$ of the pixels, are filled in from the edge inward with the local mean of the surrounding pixels plus a small random scatter. The brightest CIB pixels are zeroed in the same spirit, which is the analogue of the point-source flux cut applied to real data. Both cuts remove the extreme tails of these two fields and make training more stable, and they are cuts that a typical field-level analysis would apply.
We implement component-specific preprocessing: $\kappa$, kSZ, and tSZ maps are
zero-centered and normalized to $[-1,1]$.
For the CIB field, whose dynamic range is larger than the other fields' in the current setup, we apply a three-step transformation consisting of a signed logarithmic transform $w = \text{sign}(x) \log(|x|/\sigma + 1)$, a standardization $(w - \mu)/\sigma_t$, and a squeezing by factor $m$ to constrain values within $[-1,1]$, where $\sigma$, $\mu$, $\sigma_t$, and $m$ are field-specific parameters derived from the training dataset.
The flux cut confines the CIB of the training maps to a fixed interval, and we hold the generated maps to the same bound. This removes the rare pixels at which the inverse of the logarithmic transform diverges, and it changes the CIB power spectrum by less than one percent in every band.
Our dataset comprises 1440 training samples and 173 validation samples. All of these maps are cutouts from the same \texttt{Agora} simulation, and none of the validation maps is a repeat of a training map.

\textbf{\textit{Model Architecture and Training}}\label{sec:Architectureandtraining}
Our multi-scale wavelet generation framework employs three CFMs tailored to different decomposition levels, with detailed specifications listed in Table~\ref{tab:architecture}: a U-Net that models the $128^2$ low-frequency map, and two convolutional conditional flows that generate the detail coefficients at $128^2$ and at $256^2$.
The two conditional flows weight each wavelet band of each field in Eq.~\ref{eq:loss} by the inverse square root of its mean-squared target. Without this weighting the CIB, whose amplitude in the transformed space is by far the largest, takes $85$ to $95\%$ of the loss and the tSZ as little as $0.6\%$. We train the base flow on channels standardized to unit variance, which serves the same purpose at the coarsest scale.

\begin{wraptable}{r}{0.70\textwidth}
  \setlength{\tabcolsep}{3pt} 
  \caption{Architectures of the three flows of the Wavelet-CFM model}
  \label{tab:architecture}
  \centering
  \small
  \begin{tabular}{lccc}
    \toprule
    & Base (level-2 low) & Level-2 detail & Level-1 detail \\
    \midrule
    Resolution & $128^2$ & $128^2$ & $256^2$ \\
    In $\to$ out channels & $4\to4$ & $12{+}4 \to 12$ & $12{+}4 \to 12$ \\
    Network & U-Net & ConvNet & ConvNet \\
    Self-attention & Yes ($32^2$ bottleneck) & No & No \\
    Parameters & $1.1$M & $8.6$M & $8.6$M \\
    Conditioning & --- & low-freq concat & low-freq concat \\
    Source prior & white noise & CC $\times$ envelope & CC $\times$ envelope \\
    Loss weighting & --- & per band & per band \\
    Batch size & 720 & 720 & 128 \\
    \bottomrule
  \end{tabular}
\end{wraptable}
Training uses the AdamW optimizer with a linear warm-up and a cosine decay of the learning rate, and four GPUs per flow.  Sampling a four-field $512^2$ map takes about a third of a second on a single A100, so a suite of a thousand $9\times9$ deg$^2$ patches costs a few GPU-minutes. A new \texttt{Agora} sky instead requires a new $N$-body simulation with $3840^3$ particles in a $1\,h^{-1}$\,Gpc box, and each component is then painted onto its lightcone~\citep{yuuki}.

To test the impact of combining the wavelet decomposition with CFMs, we compare our model with two alternatives. The first is the Wavelet Flow model of~\citep{Mebratu:2025vxc}, which keeps the wavelet decomposition but uses normalizing flows instead of CFMs. The second is a single CFM trained directly on the full $512^2$ four-field maps with no wavelet decomposition. Both are trained on the identical four-field data with identical preprocessing.
Over $300 < \ell < 1000$ the power of the Wavelet Flow is low by $11\%$ in the kSZ and by $15\%$ in the tSZ, three to five times the deficit of the Wavelet-CFM chain. It also loses two thirds of the kSZ kurtosis, where the Wavelet-CFM chain loses at most a tenth.
The single CFM uses the U-Net of our base flow with the \textit{torchcfm}~\citep{tong2024improving} settings for $512^2$ images and a white noise prior. Its $19.5$M parameters roughly match the $18.3$M of our three flows together. Over $300 < \ell < 1000$ its power is low by $11\%$ in the kSZ and by $10\%$ in the tSZ, about three times the deficit of the Wavelet-CFM chain. At these scales the base flow of the Wavelet-CFM chain also starts from white noise, so the difference comes from the factorization and not from the prior. The single CFM does better than the Wavelet-CFM chain above $\ell \approx 3000$, where its tSZ power is high by $2\%$ instead of low by $5\%$.

\section{Results, Limitations, and Conclusions}
\label{sec:results}
\begin{figure}[t]
    \centering
    \begin{minipage}[t]{0.495\textwidth}
        \centering
        \includegraphics[width=\textwidth]{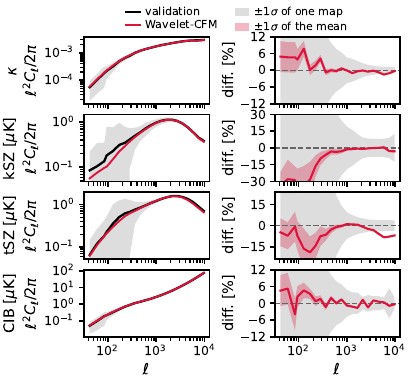}  
    \end{minipage}
    \hfill
    \begin{minipage}[t]{0.495\textwidth}
        \centering
        \includegraphics[width=\textwidth]{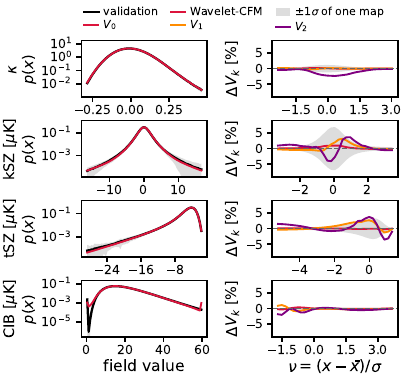}  
    \end{minipage}
    \caption{\textit{(Leftmost)} Auto power spectra of validation maps and model samples
    (173 and 2000 maps), and \textit{(Center left)} their fractional difference, for $\kappa$, kSZ, tSZ and CIB (top to
    bottom). \textit{(Center right)} One-point distributions of the same fields, and \textit{(Rightmost)} the deviation of the
    three Minkowski functionals (area $V_0$, perimeter $V_1$ and Euler characteristic $V_2$) from
    the validation maps, as a percentage of each functional's peak. Grey bands represent the
    1$\sigma$ map-to-map scatter of the validation spectra, one-point distributions, and $V_0$. Red bands are the 1$\sigma$ uncertainty of the mean fractional difference.}
    \label{fig:result}
\end{figure}

\begin{figure}[t]
    \centering
    \includegraphics[width=\textwidth]{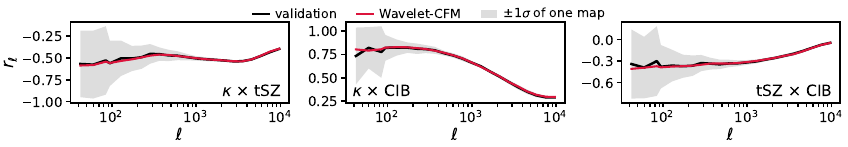} 
    \caption{Cross-correlation coefficients $r_\ell$ for the three pairs that do not involve the kSZ, from the
    same maps as Figure~\ref{fig:result}. $r_\ell$ is computed for each map and averaged over maps.
    Grey bands represent the 1$\sigma$ map-to-map scatter of the validation $r_\ell$. The three pairs involving the kSZ are consistent with zero, since the sign of
    the line-of-sight velocity is random \citep{kszcrossPhysRevLett.117.051301, 2017A&Akszcross}, and are not shown here.}
    \label{fig:cross}
\end{figure}
We trained Wavelet-CFM models on joint $\kappa$-kSZ-tSZ-CIB maps using a hybrid prior.
The left half of Figure~\ref{fig:result} presents the auto power spectra of 2000 generated maps and the 173 validation maps.

Averaged over broad bands above $\ell \approx 1000$, the $\kappa$, kSZ and CIB spectra agree with the validation maps to within $1\%$, and between $\ell \approx 300$ and $1000$ the kSZ and tSZ deficits are $4 \pm 3\%$ and $3 \pm 3\%$, both consistent with zero. The tSZ is the largest deviation, and it is low by $5\%$ above $\ell \approx 3000$ and by $8\%$ near $\ell \approx 6000$. Below $\ell \approx 300$ the kSZ and tSZ band powers are dominated by rare bright objects and 173 validation maps do not pin their mean so we make no claim there.

We similarly recover the one-point distributions and the Minkowski functionals to within a few percent of their peak amplitude in every field, as the right half of Figure~\ref{fig:result} shows, so the model is learning the non-Gaussian statistics of these fields and not just their power spectra. The clearest exception is the CIB, where the flux cut gives the data a hard ceiling and a spike of zeroed pixels that a continuous flow cannot reproduce exactly.

Figure~\ref{fig:cross} shows the cross-correlations. $\kappa\times$CIB is the strongest, reaching $|r_\ell| \approx 0.8$ on large scales, followed by $\kappa\times$tSZ and tSZ$\times$CIB. The model reproduces the three non-kSZ pairs to $|\Delta r_\ell| \le 0.07$, against a floor of $0.03$ to $0.05$ between two disjoint sets of validation-size maps.

To check that the model is not overfitting the training set, we measure how close the generated maps come to the training maps. A model that overfits reproduces training maps, so its samples sit closer to the training set than independent maps such as the validation maps. We standardize each field with its mean and standard deviation over the training set and compare all four fields at full resolution. The median distance from a generated map to its nearest training map is then $1.356$, against $1.358$ for a validation map, and the closest generated map is as far from the training set as the closest validation map. The generated maps therefore show no sign of overfitting.

This work improves on the two closest existing models: the wavelet flow of~\citep{Mebratu:2025vxc} and the denoising diffusion model of~\citep{Prabhu:2025vmu}. Each generates only two of these four fields, and on smaller maps. Before calibration the diffusion model's pixel standard deviation
is low by $12.5\%$ for the tSZ and by $3\%$ for the CIB, and it is corrected with a post-hoc rescaling; though not directly comparable due to differences between the two setups, the corresponding factors here, computed the same way, are $2$ to $3\%$ for the Sunyaev-Zel'dovich fields and below $1\%$ for $\kappa$ and CIB. %
Our reported results are not scaled.
Three limitations remain: the model does not reproduce the hard edges that the CIB flux cut imposes on the one-point distributions; it is trained on a single simulation at a single cosmology; and it is trained on only one tSZ and CIB frequency. Real maps satisfy the bound we place on the generated CIB by construction, so the likelihood the model assigns to an observation is unchanged. However, our samples no longer follow the density defined by the flow exactly, as the clip moves $0.1\%$ of the CIB pixels, and we have not quantified its effect on inference.

These results demonstrate that Wavelet-CFM models can effectively capture complex multi-field correlations and non-Gaussian statistics across multiple observational channels, with a tractable likelihood. This is an important first step towards incorporating non-Gaussian foreground fields in field-level analysis of mm-wave data~\citep{muse3g, bayesian_sptpol}.

\label{endofcontent}
\begin{ack}
{We acknowledge the following publicly available Python packages used in this work: \textit{matplotlib}~\citep{Matplotlib}, \textit{numpy}~\citep{2020NumPy-Array}, \textit{pandas}~\citep{pandas}, \textit{pixell}~\citep{2021pixell}, \textit{pytorch}~\citep{pytorch}, \textit{scipy}~\citep{2020SciPy-NMeth}, \textit{torchcfm}~\citep{tong2024improving}, and \textit{torchdiffeq}~\citep{torchdiffeq}.

This work used the resources of the SLAC Shared Science Data Facility (S3DF) at SLAC National Accelerator Laboratory. S3DF is a shared High-Performance Computing facility, operated by SLAC, that supports the scientific and data-intensive computing needs of all experimental facilities and programs of the SLAC National Accelerator Laboratory. SLAC is operated by Stanford University for the U.S. Department of Energy's Office of Science.
This research used resources of the National Energy Research Scientific Computing Center (NERSC), a Department of Energy Office of Science User Facility.

The authors acknowledge support from an Early Career Research Award DE-SC0026376 of the Department of Energy.}
\end{ack}

\newpage
{\small
\bibliographystyle{unsrtnat}
\bibliography{wavelet_foregrounds}}

\newpage
\appendix
\section{Log-likelihood on Gaussian maps}
\label{app:gauss}
To test the log-likelihood we trained the same model on Gaussian maps whose density is known exactly. We drew $10^4$ training maps and $10^3$ validation maps of $\kappa$ and CIB with $128^2$ pixels at 1-arcminute resolution. The maps are periodic Gaussian random fields with the same means and the same auto and cross power spectra as the training maps of Section~\ref{sec:methods}. We preprocess them in the same way as in the main text, but we recompute the constants of the $\kappa$ and CIB transforms from these maps. We use a single wavelet level, so a U-Net with a white noise prior models the $64^2$ low-frequency map and a conditional flow with the CC $\times$ envelope prior models the detail coefficients. The log-likelihood of a map is the sum of the two flow terms and the log-Jacobian of the preprocessing in Eq.~\ref{eq:logp}, and we compare it with the exact Gaussian log-density of the same map.
\begin{equation}
\label{eq:logp}
\begin{split}
\log p(x) = {} & \log p_0(z_0) - \int_0^1 \nabla \cdot v^{\rm b}_\theta(z_t, t)\, dt + N_y \sum_c \log f_c \\
& + \log q_0(h_0 \mid y) - \int_0^1 \nabla \cdot v^{\rm d}_\theta(h_t, t \mid y)\, dt + \log \left| \det \frac{\partial D}{\partial x} \right|
\end{split}
\end{equation}%
Here $D$ is the preprocessed map, $y$ its $64^2$ low-frequency map with $N_y$ pixels per field, and $h$ its detail coefficients. The path $z_t$ of the base flow ends at $z_1 = f \odot y$ at $t=1$, where $f_c$ scales field $c$ to unit variance. The path $h_t$ of the detail flow ends at $h_1 = h$. We obtain both paths by integrating the flows backward from $t=1$ to $t=0$. We estimate each divergence with the Hutchinson trace estimator, $\nabla \cdot v_\theta \approx \frac{1}{K} \sum_k \epsilon_k^{\top} (\partial v_\theta / \partial u)\, \epsilon_k$, where $u$ stands for $z_t$ or $h_t$ and the $\epsilon_k$ are random vectors of $\pm 1$ entries. $p_0$ is the white noise prior and $q_0(h_0 \mid y) = \mathcal{N}(h_0 / s(y); 0, \Sigma_{\rm CC}) / \prod s(y)$ is the CC $\times$ envelope prior with envelope $s(y)$ and CC covariance $\Sigma_{\rm CC}$. The Haar transform is orthonormal and adds no term. The preprocessing acts on each pixel separately, so its log-Jacobian is a sum over pixels.

\begin{figure}[h]
    \centering
    \includegraphics[width=0.8\textwidth]{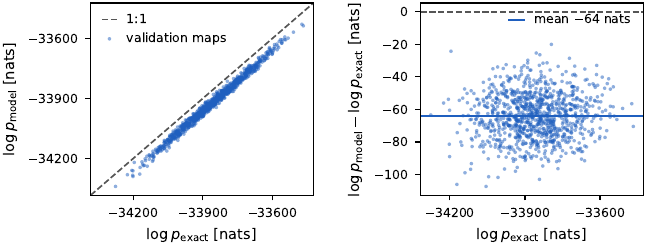}
    \caption{Log-likelihood of the 1000 Gaussian validation maps under the model against their exact log-density (left), and the difference between the two (right). The dashed lines mark perfect agreement.}
    \label{fig:gauss}
\end{figure}

We evaluate the log-likelihood on all 1000 validation maps with $K = 32$ probes and the dopri5 solver. We use the base flow at $70{,}000$ steps and the detail flow at $65{,}000$ steps. After that the detail flow starts to overfit. Figure~\ref{fig:gauss} shows that the model follows the exact log-density from map to map with a slope of $1.007$. On average it is $64$ nats (or $0.002$ nats per pixel and field) below the exact log-density.

\end{document}